\documentclass[12pt]{iopart}
\pdfoutput=1
\usepackage[utf8]{inputenc} 			
\usepackage[T1]{fontenc}				
\usepackage{lmodern}					
\usepackage[pdftex]{graphicx} 		    
\usepackage[svgnames]{xcolor}
\usepackage{cite}
\usepackage{iopams}                     
\usepackage{setstack}                   
\usepackage[english]{babel}
\usepackage{stmaryrd}
\usepackage{amssymb}
\usepackage{bbm}
\usepackage{caption}

   \let\P=\Pi

\def\ee{e}
    
\def\to{\rightarrow}

\begin{document}

\title{Nonlinear analog of the complexity-stability transition in random dynamical systems:
a replica calculation}
\author{Thibaut Arnoulx de Pirey$^{1,2}$, Frédéric van Wijland$^{1}$}
\address{$^1$ Laboratoire Matière et Systèmes Complexes, Université Paris Cité \& CNRS (UMR 7057), 10 rue Alice Domon et Léonie Duquet, 75013 Paris, France}
\address{$^2$ Department of Physics, Technion-Israel Institute of Technology, Haifa 32000, Israel}
\ead{thibaut.arnoulxdepirey@univ-paris-diderot.fr}
\date{\today\ -- \jobname}

\begin{abstract}
We consider large-dimensional dynamical systems involving a linear force and a random force comprising both potential and non-conservative contributions. Such systems  are known to exhibit a topological trivialization phase transition as the strength of the random force is increased. This is reflected in the number of stationary points of the dynamical systems that transitions from one to an exponential in the number of degrees of freedom. We analyze this transition by means of a replica calculation.

\end{abstract}

\section{Introduction}
The question of how to describe the dynamics of a large complex system was initially investigated, at the linear level, by Ashby and Gardner \cite{gardner1970connectance} and subsequently formalized in \cite{somorjai1972relationship,cohenpb1985will,may1972will} by  encapsulating their complexity in terms of random stability matrices. Beyond these linear descriptions, in a recent work~\cite{fyodorov2016nonlinear}, Fyodorov and Khoruzhenko asked about the 
number $\mathcal N$ of stationary points an $N$-dimensional dynamical system with random forces possesses in the large $N$ limit. To be specific, they considered $N$ degrees of freedom $x_i$ for $i=1,\dots,N$ evolving according to
\begin{eqnarray}\label{eq:EOMfyod}
  \frac{{\rm d} x_i}{{\rm d} t}  = - \mu x_i + f_i(x_1,\dots,x_N) \, ,
\end{eqnarray}
where the force field $\bi{f}$ involves both gradient and solenoidal random contributions,
\begin{eqnarray}
  f_i(\bi{x}) = -\partial_{x_i}V(\bi{x}) + \frac{1}{\sqrt{N}}\sum_{j = 1}^{N}\partial_{x_j}A_{ij}(\bi{x}) \, ,
\end{eqnarray}
with $A_{ij}$ antisymmetric for the associated contribution in the equation of motion to be divergence free. Both $V(\bi{x})$ and $A_{ij}(\bi{x})$ are chosen to be independent Gaussian zero mean random fields with correlations
\begin{eqnarray}\label{eq:correlfunc}
        \left\langle V(\bi{x})V(\bi{y}) \right\rangle = v^2 \Gamma_V\left(\left|\bi{x}-\bi{y}\right|^2\right) \, \, , \, \, \Gamma_V''(0) = 1 \, , \\
        \left\langle A_{ij}(\bi{x})A_{mn}(\bi{y}) \right\rangle = a^2 \Gamma_A\left(\left|\bi{x}-\bi{y}\right|^2\right)\left(\delta_{im}\delta_{jn}-\delta_{in}\delta_{jm}\right) \, \, , \, \, \Gamma_A''(0) = 1 \, ,
\end{eqnarray}
The motivation of Fyodorov and Khoruzhenko was to probe the extent to which the instability mechanism brought forth by May~\cite{may1972will}, within the context of the population dynamics of large ecosystems, extends to nonlinear dynamical systems. They obtain an explicit expression for the average number of critical (or stationary) points $\langle{\mathcal N}\rangle$ where $-\mu\bi{x}+\bi{f}(\bi{x})=\mathbf{0}$. Our goal in this technical note is to show how the log asymptotic of their remarkable result can be recovered without resorting to the theory of random matrices, using a replica approach \cite{dotsenko2005introduction}. The latter, unlike the elegant random matrix approach used in \cite{fyodorov2016nonlinear}, comes with its fair share of conceptual difficulties and uncontrolled mathematics (and to begin with, it is based on a complex analysis theorem~\cite{hardy1920two} the hypotheses of which are never checked). However it is technically not only somewhat simpler, but also perhaps closer to techniques more commonly used in theoretical physics. Furthermore, despite all the possible caveats behind it, the replica method is often the only one at hand. In the few instances where exact results are available, it is therefore interesting to use them as a benchmark to test the predictions of the replica approach. This is why we believe a simple alternative proof of the results of \cite{fyodorov2016nonlinear} deserves to be presented independently as a complement.

\section{Derivation}
\subsection{Setting the stage, and the goal}
Sticking to the notations of \cite{fyodorov2016nonlinear} we introduce the parameter $\tau = \frac{v^2}{v^2+a^2}$ that measures the relative strength of the gradient and solenoidal terms. We also define $m = \frac{\mu}{2\sqrt{v^2+a^2}\sqrt{N}}$ that compares the amplitude of the linear contribution in Eq.~\eref{eq:EOMfyod} to the nonlinear ones. The main result of \cite{fyodorov2016nonlinear} is that $\left\langle \mathcal{N} \right\rangle$, the mean number of stationary points of the dynamical system, undergoes at large $N$ a phase transition from a regime where it is $O(1)$ at small $m$, \textit{i.e.} a regime where the harmonic potential dominates the dynamics, to a regime where $\langle\mathcal{N}\rangle$ scales exponentially with the system size $N$. Following \cite{fyodorov2016nonlinear}, the mean number of stationary points of Eq.~\eref{eq:EOMfyod} can be obtained from the Kac-Rice formula for the number $\mathcal{N}$ of stationary points in a given realization of Eq.~\eref{eq:EOMfyod},
\begin{eqnarray} 
     \mathcal{N} = \int d\bi{x}\, \delta\left(-\mu\bi{x}+\bi{f}(\bi{x})\right)\left|\det\left(-\mu \mathbbm{1} + \bi{\partial}\bi{f}(\bi{x})\right)\right| .
\end{eqnarray}
The correlation functions in Eq.~\eref{eq:correlfunc} being smooth functions of the distance square, $\left\langle f_i(\bi{x})\partial_j f_\ell(\bi{x}) \right\rangle = 0$ and thus the determinant of the Jacobian, the statistics of which are translationnally invariant, can be pulled outside the integral sign after averaging. After little algebra, this leads to 
\begin{eqnarray} \label{eq:abs_value}
    \left\langle \mathcal{N} \right\rangle = \left\langle \left| \det{\left(\delta_{ij} + \frac{J_{ij}}{\mu}\right)}\right| \right\rangle \,,
\end{eqnarray}
with $J_{ij} \equiv \partial_i f_j(0)$. The $J_{ij}$'s are the coefficients of an $N \times N$ zero mean Gaussian matrix with correlations
\begin{equation}\label{eq:corrlJ}
\left\langle J_{ij} J_{kl} \right\rangle = \alpha^2\left( \left(1+\frac{1-\tau}{N}\right) \delta_{ik}\delta_{jl} + \left(\tau - \frac{1-\tau}{N}\right)\left( \delta_{ij}\delta_{kl} + \delta_{il}\delta_{jk}\right)\right) \, ,
\end{equation}
and with $\alpha = 2\sqrt{v^2+a^2}$. We compute $\langle{\mathcal N}\rangle$ in Eq.~\eref{eq:abs_value} by means of a replica calculation of the absolute value of a determinant. It has been shown in \cite{BenArousFyodorov} that, in the multiple equilibria phase, \textit{i.e.} for $m < 1$, the annealed complexity of stable stationary points of Eq.~\eref{eq:EOMfyod} is strictly smaller than that of stationary points irrespective of their index. Therefore, the absolute value cannot be neglected when performing the average in Eq.~\eref{eq:abs_value}. Different strategies have been put forward in the literature to compute the mean of the absolute value of a random matrix determinant \cite{fyodorov2004complexity, kurchan1991replica}. In this work, we use the following identity valid for any real matrix
\begin{eqnarray}
    \left|\det{\left(\mathbbm{1} + J/\mu\right)}\right| = \lim_{\epsilon \to 0^+} \lim_{n \to 0} I_{\epsilon}^{n-1} \, ,
\end{eqnarray}
with
\begin{eqnarray}
    I_{\epsilon} = \frac{1}{\sqrt{\det\left(\left(\mathbbm{1} + J/\mu\right)\left(\mathbbm{1} + J/\mu\right)^T + \epsilon^2 \mathbbm{1}\right)}} \,, \nonumber \\ \hphantom{I_{\epsilon}} = \left(\det\left[ 
    \begin{array}{cc} 
      \epsilon & i\left(\mathbbm{1} + J/\mu\right) \\ 
      \vspace{-0.55cm} \\
      i\left(\mathbbm{1} + J/\mu\right)^T & \epsilon
    \end{array} \right]\right)^{-\frac{1}{2}} \,, \nonumber \\ \hphantom{I_{\epsilon}} = \int \prod_{i=1}^N \frac{d \phi_i d \varphi_i}{2\pi} \ee^{-\frac{\epsilon}{2}\bphi^2-\frac{\epsilon}{2}\bvarphi^2 + i \sum_{i,j = 1}^N \phi_i \left(\delta_{ij} + J_{ij}/\mu\right)\varphi_j} \, .
\end{eqnarray}
The parameter $\epsilon > 0$ is introduced to guarantee the convergence of the above integral. We therefore obtain, 
\begin{eqnarray}
    \left\langle \mathcal{N} \right\rangle = \lim_{\epsilon \to 0^+} \lim_{n \to 0} \left\langle I_{\epsilon}^{n-1} \right\rangle \, .
\end{eqnarray}
The average $\left\langle I_{\epsilon}^{n-1} \right\rangle$ is then computed for $n \in \mathbb{N}$ with $n > 1$ and $\left\langle \mathcal{N} \right\rangle$ is obtained using an analytical continuation to $n = 0$. The validity of this analytical continuation, guaranteed when the hypotheses of Carlson's theorem are verified, will not be checked here. Furthermore, as is usual in the use of the replica trick, we assume that the limits $\lim_{N \to \infty}$ and $\lim_{\epsilon \to 0^+}\lim_{n \to 0}$ commute. We will critically discuss this assumption at the end. Hereafter, we use $a,b$ to label the different replicas. Introducing the overlaps between the different replicated fields,
\begin{eqnarray}
        P_{ab} = \frac{1}{N}\sum_{i}\varphi_i^a \varphi_i^b \, , \\
        Q_{ab} = \frac{1}{N}\sum_{i}\phi_i^a \phi_i^b \, , \\
        R_{ab} = \frac{1}{N}\sum_{i}\varphi_i^a \phi_i^b \, ,
\end{eqnarray}
we obtain
\begin{eqnarray}\label{eq:integral}
\fl \left\langle I_{\epsilon}^{n-1} \right\rangle = \int \prod_{i = 1}^N \prod_{a = 1}^{n-1}\frac{d \phi_i^a d \varphi_i^a}{2\pi} \exp\left(-\frac{1-\tau}{2m^2}\left[{\rm Tr}\left(QP\right) - \left({\rm Tr}R^2 + \left({\rm Tr}R\right)^2\right)\right]\right) \nonumber \\ \!\!\!\!\! \!\!\!\!\! \!\!\!\!\! \!\!\!\!\! \exp\left\{N \left[-\frac{\epsilon}{2}{\rm Tr}Q - \frac{\epsilon}{2}{\rm Tr}P + i {\rm Tr}R - \frac{1}{2 m^2}\left( {\rm Tr}\left(QP\right) + \tau {\rm Tr}R^2 + \tau\left({\rm Tr}R\right)^2\right)\right]\right\} \,. 
\end{eqnarray}
By now using the overlaps as new integration variables~\cite{fyodorov2007classical}, the above equation can be rewritten in a form suitable for saddle point evaluation in the large $N$ limit
\begin{eqnarray}\label{eq:integral2}\fl
\left\langle I_{\epsilon}^{n-1} \right\rangle = C_{N,n} \int_{S > 0} \prod_{a \leq b}^{n-1} d Q_{ab} d P_{ab} \prod_{a, b}^{n-1} d R_{ab} \left(\det S\right)^{-n + \frac{1}{2}} \rme^{N f(S)}\nonumber \\ \exp\left(-\frac{1-\tau}{2m^2}\left[{\rm Tr}\left(QP\right) - \left({\rm Tr}R^2 + \left({\rm Tr}R\right)^2\right)\right]\right)
\end{eqnarray}
where $S$ is the $2(n-1) \times 2(n-1)$ symmetric matrix of overlaps with block entries given by
\begin{eqnarray}
S = \left[ 
    \begin{array}{c|c} 
      P & R \\ 
      \hline \vspace{-0.55cm} \\
      R^T & Q
    \end{array} 
    \right] \, ,
\end{eqnarray}
and the function $f$ is defined by
\begin{eqnarray}
\fl f(S) = -\frac{\epsilon}{2}{\rm Tr}Q - \frac{\epsilon}{2}{\rm Tr}P + i {\rm Tr}R - \frac{1}{2 m^2}\left( {\rm Tr}\left(QP\right) + \tau {\rm Tr}R^2 + \tau\left({\rm Tr}R\right)^2\right) \nonumber \\ + (n-1) + \frac{1}{2}\ln \det S \,.
\end{eqnarray}
The integration domain in Eq.~\eref{eq:integral2} is restricted to positive definite $S$ matrices. Lastly, in the limit $N \to \infty$, the constant $C_{N,n}$ arising from the Jacobian of the transformation from the fields to the overlaps reads
\begin{eqnarray}
  C_{N,n} \underset{N \to \infty}{\sim} \left(\frac{1}{2}\right)^{n-1} \left(\frac{N}{2\pi}\right)^\frac{(n-1)(2n-1)}{2} \,.
\end{eqnarray}
The number of integration variables being now independent of $N$, the integral in Eq.~\eref{eq:integral2} can be evaluated in the large $N$ limit by means of a saddle point approximation. We introduce for the inverse $S^{-1}$ the block notation
\begin{eqnarray}\label{eq:inverse_notation}
S^{-1} = \left[ 
    \begin{array}{c|c} 
      \tilde{P} & \tilde{R} \\ 
      \hline \vspace{-0.45cm}  \\
      \tilde{R}^T & \tilde{Q}
    \end{array} 
    \right] \, .
\end{eqnarray}
The saddle point equations associated to Eq.~\eref{eq:integral} then read
\begin{eqnarray}\label{eq:saddle_point}
      \tilde{P}_{ab} - \frac{1}{m^2}Q_{ab} - \epsilon \delta_{ab} = 0 \, , \nonumber \\
      \tilde{Q}_{ab} - \frac{1}{m^2}P_{ab} - \epsilon \delta_{ab} = 0 \, , \nonumber \\
      \tilde{R}_{ab} + i \delta_{ab} - \frac{\tau}{ m^2}\left( \delta_{ab} {\rm Tr}R + R_{ba}\right) = 0 \, .
\end{eqnarray}
Our task is now to solve this set of equations.

\subsection{Block identity ansatz}\label{sec:diago}
We look for a solution of the saddle point equations \eref{eq:saddle_point} in the form of a block identity matrix
\begin{eqnarray}
S = \left[ 
    \begin{array}{c|c} 
      p\mathbbm{1} & r \mathbbm{1} \\ 
      \hline 
      r \mathbbm{1} & q \mathbbm{1}
    \end{array} 
    \right] \, .
\end{eqnarray}
This ansatz will be critically discussed in section \ref{sec:ansatz}. The equation Eq.~\eref{eq:saddle_point} then reduces to
\begin{eqnarray} \label{eq:spdiag}
       \frac{q}{pq-r^2}-\frac{q}{m^2} - \epsilon = 0 \, ,  \nonumber \\
       \frac{p}{pq-r^2}-\frac{p}{m^2} - \epsilon = 0 \, , \nonumber \\
       \frac{r}{pq-r^2} - i + \frac{n \tau}{m^2}r = 0 \, .
\end{eqnarray}
Interestingly, in the limit $ n \to 0$, the parameter $\tau$ disappears from the saddle point equations. There exists three triplets of solutions, each of them with $p = q$.  As $\epsilon \to 0$, the solutions to the $n=0$ saddle point equations read
\begin{eqnarray}\label{eq:sol}
    p = q = 0, \, r = i \, , \nonumber \\
    p = q = \pm \sqrt{m^2 - m^4}, \,  r = i m^2 \, .
\end{eqnarray}
At exactly $\epsilon = 0$, there is a degeneracy of the latter solution along the hyperbola $pq = m^2(1-m^2)$. Lifting this degeneracy is the purpose of the small $\epsilon$ parameter. As $n \to 0$ and $\epsilon \to 0$ the exponential weight evaluated at the saddle point is given by
\begin{eqnarray}
    f(p = q = 0,r = i) = 0 \, , \nonumber \\
    f(p = q = \pm \sqrt{m^2 - m^4}, r = i m^2) = \frac{m^2-1}{2} - \ln m \, . 
\end{eqnarray}
Note that the dependence on $\tau$ is completely washed out, a remarkable feature already noted in \cite{fyodorov2016nonlinear}.

\subsection{Saddle point selection}
In this section, we show that for $m > 1$ the first solution is selected while the second one is selected for $m < 1$. We start by expanding $P,Q,R$ around diagonal matrices as 
\begin{eqnarray}
P_{ab} = p_a \delta_{ab} + \frac{1}{\sqrt{N}}\delta P_{ab}\left(1-\delta_{ab}\right) \, , \nonumber \\
Q_{ab} = q_a \delta_{ab} + \frac{1}{\sqrt{N}}\delta Q_{ab}\left(1-\delta_{ab}\right) \, , \nonumber \\
R_{ab} = r_a \delta_{ab} + \frac{1}{\sqrt{N}}\delta R_{ab}\left(1-\delta_{ab}\right) \, .
\end{eqnarray}
Here we disregard the subexponential multiplicative constants that will be carefully dealt with in section \ref{sec:fluc_det} and we use the $\sim$ sign to express a log equivalence. Upon linearizing the $\left({\rm Tr}R\right)^2$ term  by use of an additional Gaussian variable, the integral in Eq.~\eref{eq:integral2} factorizes and reads 
\begin{eqnarray}\label{eq:integral3}
\fl \left\langle I_{\epsilon}^{n-1} \right\rangle \sim \int_{-\infty}^{+\infty} d z \, {\rm exp}\left(-N\frac{z^2}{2}\right)\left(\int_0^{+\infty} d q d p \int_{-\sqrt{pq}}^{\sqrt{pq}}  d r \exp{\left[N \left(-\frac{\epsilon}{2}(p+q) + i(1+\frac{\sqrt{\tau}}{m}z)r \right.\right.}\right. \nonumber \\ \left.\left.\left. -\frac{1}{2m^2}\left(pq + \tau r^2\right) + 1 + \frac{1}{2}\ln\left(pq -r^2\right)\right)\right]\right)^{n-1} \,.
\end{eqnarray}
As far as the leading exponential behavior is concerned, $\left\langle I_{\epsilon}^{n-1} \right\rangle$ can thus be obtained as
\begin{eqnarray}\label{eq:integral4}
\fl \left\langle I_{\epsilon}^{n-1} \right\rangle \sim \int_{-\infty}^{+\infty} d z \, {\rm exp}\left(-N\frac{z^2}{2}\right)\int_0^{+\infty} d q d p \int_{-\sqrt{pq}}^{\sqrt{pq}}  \rmd r \exp\left[N(n-1) \left(-\frac{\epsilon}{2}(p+q)\right.\right. \nonumber \\ \left.\left. + i(1+\frac{\sqrt{\tau}}{m}z)r  -\frac{1}{2m^2}\left(pq + \tau r^2\right) + 1 + \frac{1}{2}\ln\left(pq -r^2\right)\right)\right]\,.
\end{eqnarray}
Introducing $u = \sqrt{pq}$ and $v = p + q$ and integrating over $v$ and $z$, we are left with the two-dimensional integral
\begin{eqnarray}
        \left\langle I_{\epsilon}^{n-1} \right\rangle & \underset{N \to \infty}{\sim} \int_{0}^{+\infty} d u \int_{-u}^{u} d r \,\, \ee^{N(n-1)g(u,r)} \, ,
\end{eqnarray}
with 
\begin{eqnarray}
    g(u,r) & = 1 + \frac{1}{2}\ln(u^2-r^2) + i r -\frac{1}{2m^2}\left(u^2 + n \tau r^2\right) \, , \\
    & \hspace{-0.15cm} \underset{n \to 0}{=} \!\! 1 + \frac{1}{2}\ln(u^2 - r^2) + ir - \frac{u^2}{2m^2} \,.
\end{eqnarray}
Lastly, in order to carry the contour deformation necessary to the saddle point evaluation, we change variables and introduce
\begin{eqnarray}
    u = x \cosh{\theta}\, , \\ r = x \sinh{\theta} \, ,
\end{eqnarray}
so that
\begin{eqnarray} \label{eq:integraltheta}
\fl  \left\langle I_{\epsilon}^{n-1} \right\rangle \underset{N \to \infty}{\sim} \int_{0}^{+\infty} d x \int_{-\infty}^{\infty} d \theta \,\, \exp{\left[N(n-1)\left(1 + \ln x + i x \sinh{\theta} - \frac{x^2}{2 m^2}\cosh^2\theta\right)\right]} \, .
\end{eqnarray}
Note that in Eq.~\eref{eq:integraltheta}, for the sake of simplicity of the expressions, we have already anticipated the $n \to 0$ limit in the function $g$ but that, in order to get the proper analytical continuation to $n = 0$, we keep working with $N(n - 1) > 0$. The saddle points of the $\theta$ integral in Eq.~\eref{eq:integraltheta} are given by
\begin{eqnarray}
    \theta = i\left(\frac{\pi}{2}+s\pi\right), \, s \in \mathbb{N}
\end{eqnarray}
or
\begin{eqnarray}
        \theta = i \arcsin{\left(\frac{m^2}{x}\right)} + 2 i s \pi, \, s \in \mathbbm{N} , \mbox{ if} \,\,\, \frac{m^2}{x} < 1 \, , \mbox{ or} \nonumber \\  
        \theta = i \left(\pi - \arcsin{\left(\frac{m^2}{x}\right)}\right) + 2 i s \pi, \, s \in \mathbbm{N} , \, \, \mbox{ if}  \,\,\,\frac{m^2}{x} < 1 \, , \, \mbox{ or} \\ 
        \theta = \pm \mbox{ arccosh}\left(\frac{m^2}{x}\right) + i \left(\frac{\pi}{2}+s\pi\right), \, s \in \mathbbm{N} , \, \, \mbox{ if}  \,\,\, \frac{m^2}{x} > 1 \, . \nonumber
\end{eqnarray}

\begin{figure}[htp]
    \begin{minipage}{0.45\textwidth}
        \centering
        \includegraphics[scale=0.3]{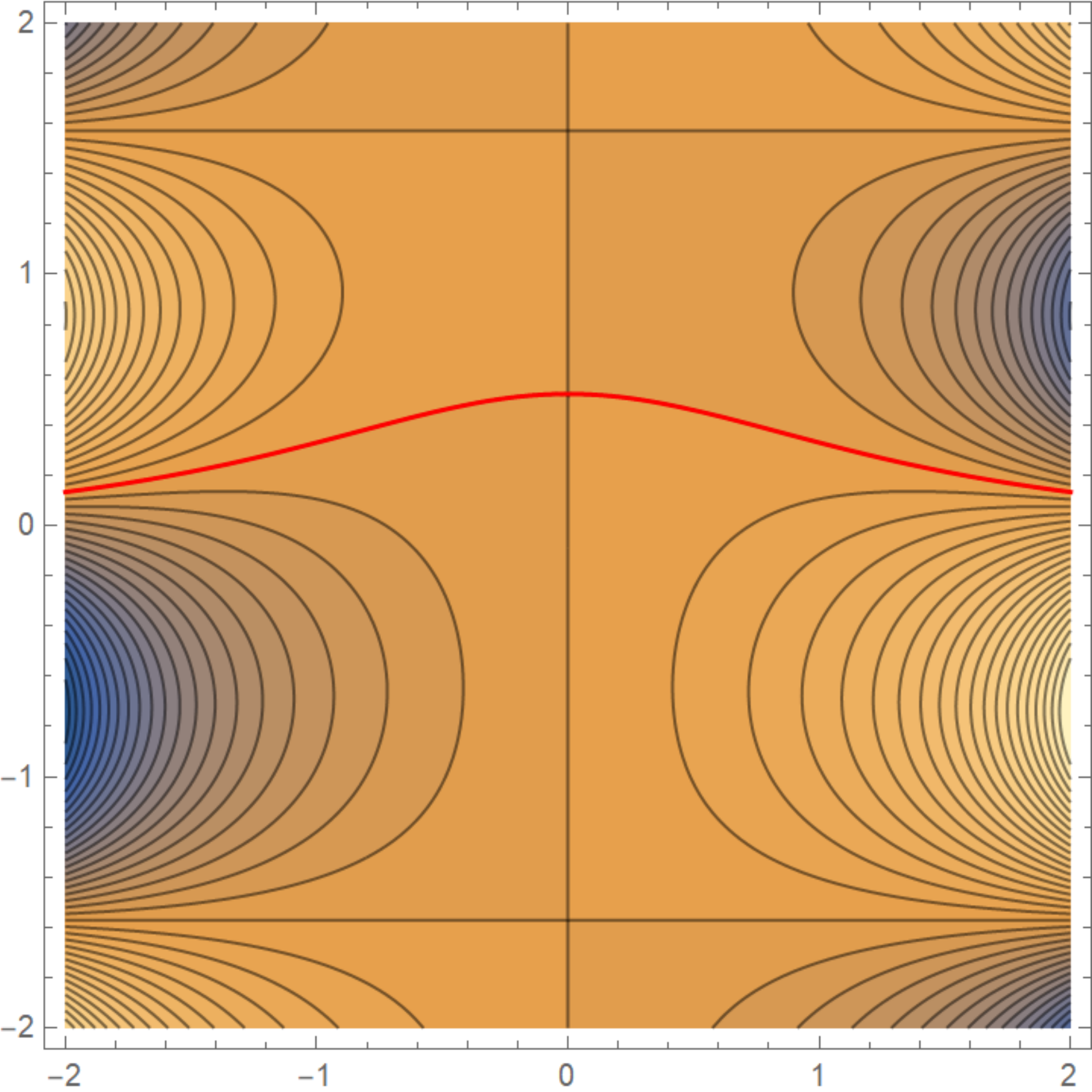} \caption{Isolines of ${\rm Im}(g)$ for \\ $m^2/x < 1$ ($m^2/x = 0.5$). In red is the deformed contour used in the saddle point approximation of \eref{eq:integraltheta}.}
        \label{fig:fig1}
    \end{minipage}
{\hfill\color{white}\vrule\hfill}
    \begin{minipage}{0.45\textwidth}
        \centering
        \includegraphics[scale=0.3]{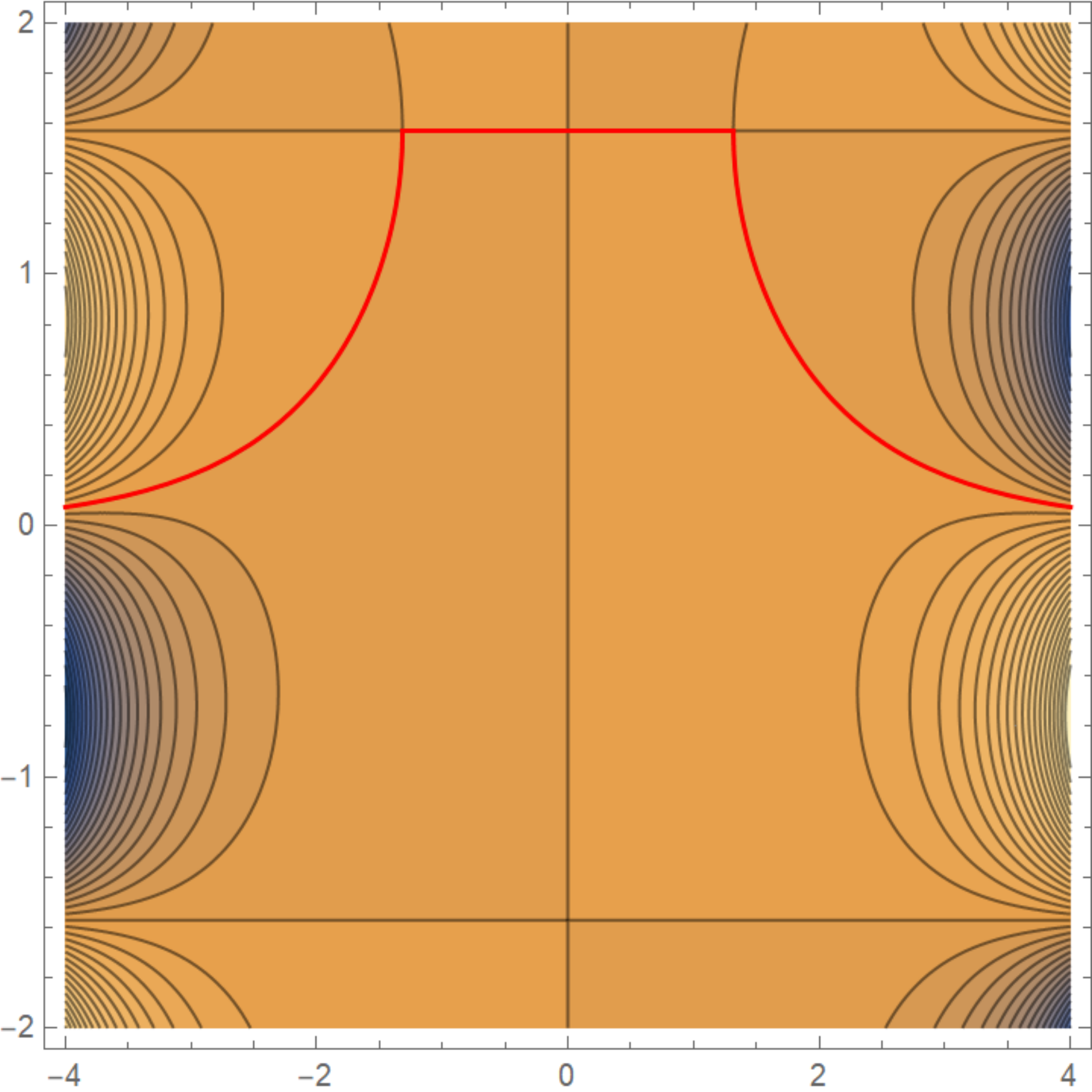}\caption{Isolines of ${\rm Im}(g)$ for \\  $m^2/x > 1$ ($m^2/x = 2$). In red is the deformed contour used in the saddle point approximation \eref{eq:integraltheta}.}
        \label{fig:fig2}
    \end{minipage}
\end{figure}

In both cases, the $\theta$ integration path is deformed to the steepest descent path of $0$ imaginary part passing through the saddles that can be seen in red respectively in Fig.~\ref{fig:fig1} and Fig.~\ref{fig:fig2}. Note that the saddle at $\theta = i \pi/2$ is only attained for $x < m^2$. Summing over all the different saddle point contributions, we therefore obtain
\begin{eqnarray}
\fl\left\langle I_{\epsilon}^{n-1} \right\rangle \underset{N \to \infty}{\sim} & \int_{m^2}^{+\infty}\!\!d x \, \exp{\left[N(n-1)\left(1 - \frac{m^2}{2} + \ln x - \frac{x^2}{2 m^2}\right)\right]} \\ + & \int_{0}^{m^2} \!\! d x \exp{\Bigg[N(n-1)\left(1 + \ln x - x\right)\!\Bigg]}  \\ + & \int_{0}^{m^2}\!\!d x \, \exp{\left[N(n-1)\left(1 - \frac{m^2}{2} + \ln x - \frac{x^2}{2 m^2}\right)\right]} \, . 
\end{eqnarray}
Therefore, if $m > 1$, the result is dominated by the second integral and we have 
\begin{eqnarray}
    \left\langle \mathcal{N} \right\rangle \sim 1 \, .
\end{eqnarray}
However if $m < 1$, the result is dominated by the first integral and we get, after taking the $n \to 0$ limit,
\begin{eqnarray}
    \left\langle \mathcal{N} \right\rangle \sim \exp{\left(-N\left(\frac{1 - m^2}{2} + \ln m\right)\right)} \, .
\end{eqnarray}
We have thus recovered, within our ansatz, the main result of \cite{fyodorov2016nonlinear} which shows a transition in the mean number of stationary points from a regime where it is $O(1)$ at $m < 1$ to a regime where it scales exponentially with the system size at $m > 1$. We stress that at the exponential level $\mathcal{N}$ does not depend on $\tau$, \textit{i.e.} on the way the non-linearities are distributed between the solenoidal and potential contributions. We now compute the contribution arising from integrating out fluctuations around our saddle point solutions.

\subsection{Multiplicative constants}\label{sec:fluc_det}
In this section, we evaluate the contributions arising from the quadratic fluctuations around the two saddle points discussed in section \ref{sec:diago}. In what proceed, $\epsilon$ is kept finite and is sent to 0 at the end. In Eq.~\eref{eq:integral2}, we expand 
\begin{eqnarray}
P_{ab} = p \delta_{ab} + \frac{1}{\sqrt{N}}\delta P_{ab} \, , \nonumber \\
Q_{ab} = q \delta_{ab} + \frac{1}{\sqrt{N}}\delta Q_{ab} \, , \nonumber \\
R_{ab} = r \delta_{ab} + \frac{1}{\sqrt{N}}\delta R_{ab} \, .
\end{eqnarray}
with $p,q,r$ the solutions of Eq.~\eref{eq:spdiag}. We denote $S_*$ the corresponding saddle point matrix and introduce $\tilde{p}, \tilde{q}, \tilde{r}$ such that
\begin{eqnarray}
S_*^{-1} = \left[ 
    \begin{array}{c|c} 
      \tilde{p}\mathbbm{1} & \tilde{r} \mathbbm{1} \\ 
      \hline 
      \tilde{r} \mathbbm{1} & \tilde{q} \mathbbm{1}
    \end{array} 
    \right] \, .
\end{eqnarray}
Using the identity,
\begin{eqnarray}
\det \left(1 + \epsilon H\right) = \exp\left(\epsilon {\rm Tr}H - \frac{\epsilon^2}{2}{\rm Tr}\left(H^2\right)\right) + O(\epsilon^3) \, ,
\end{eqnarray}
and keeping track of quadratic fluctuations only, we obtain 
\begin{eqnarray}
\fl
N f(S) = N f(S_*) - \frac{1}{2m^2}\left( {\rm Tr}\delta P \delta Q + \tau  {\rm Tr}\delta R^2 + \tau \left({\rm Tr}\delta R\right)^2 \right) - \frac{1}{4}{\rm Tr}\left(\left(S_*^{-1}\delta S\right)^2\right) \nonumber \, . \\ \fl = N f(S_*) - \frac{\tilde{p}^2}{4}{\rm Tr}\delta P^2 - \frac{\tilde{p}^2}{4}{\rm Tr}\delta Q^2 - \left(\frac{1}{2 m^2} + \frac{\tilde{r}^2}{2}\right){\rm Tr}\delta Q \delta P - \left(\frac{\tau}{2 m^2} + \frac{\tilde{r}^2}{2}\right){\rm Tr}\delta R^2 \nonumber \\ \fl  - \frac{\tilde{p}^2}{2}{\rm Tr}\delta R\delta R^T - \tilde{p}\tilde{r} \left({\rm Tr}\delta R\delta P + {\rm Tr}\delta R\delta Q\right) - \frac{\tau}{2m^2}\left( {\rm Tr}\delta R\right)^2 \, .
\end{eqnarray} 
Using a Hubbard-Stratonovich transformation to linearize the $\left( {\rm Tr}\delta R\right)^2$ term  and then splitting the above expression between diagonal and off-diagonal contributions, we get
\begin{eqnarray}
\fl
\left\langle I_{\epsilon}^{n-1} \right\rangle = \left(\frac{1}{2}\right)^{n-1} \left(\det S_*\right)^{-n+ \frac{1}{2}} \rme^{N f(S_*)} \left(\det M_1\right)^{-\frac{(n-1)(n-2)}{4}} \exp\left(\frac{(1-n)(1-\tau)}{2m^2}\left[p^2-n r^2\right]\right)\nonumber \\  \sqrt{\frac{m^2}{\tau}} \int \frac{\rmd z}{\sqrt{2\pi}} \rme^{-\frac{z^2 m^2}{2\tau}} \left(\left(\det M_2\right)^{-\frac{1}{2}} \exp{\left(- \frac{z^2}{2}(M_2^{-1})_{33}\right)} \right)^{n-1}  \, .
\end{eqnarray}
with 
\begin{eqnarray}
M_1 = \left[ 
    \begin{array}{c|c|c|c} 
      \tilde{p}^2 & \frac{1}{m^2} + \tilde{r}^2 & \tilde{p}\tilde{r} & \tilde{p}\tilde{r} \\ 
      \hline 
      \frac{1}{m^2} + \tilde{r}^2 & \tilde{p}^2 & \tilde{p}\tilde{r} & \tilde{p}\tilde{r} \\
      \hline
      \tilde{p}\tilde{r} & \tilde{p}\tilde{r} & \tilde{p}^2 & \frac{\tau}{m^2} + \tilde{r}^2 \\
      \hline 
       \tilde{p}\tilde{r} & \tilde{p}\tilde{r} & \frac{\tau}{m^2} + \tilde{r}^2 & \tilde{p}^2 
    \end{array} 
    \right] \, , 
\end{eqnarray}
and
\begin{eqnarray}
M_2 = \left[ 
    \begin{array}{c|c|c} 
      \frac{\tilde{p}^2}{2} & \frac{1}{2m^2} + \frac{\tilde{r}^2}{2} & \tilde{p}\tilde{r} \\ 
      \hline 
      \frac{1}{2m^2} + \frac{\tilde{r}^2}{2}  & \frac{\tilde{p}^2}{2} & \tilde{p}\tilde{r} \\
      \hline
      \tilde{p}\tilde{r} & \tilde{p}\tilde{r} & \tilde{p}^2 + \tilde{r}^2 + \frac{\tau}{m^2}
    \end{array} 
    \right] \, .
\end{eqnarray}
The matrix $M_1$ quantifies the fluctuations of the non block diagonal terms around $0$ while the matrix $M_2$ quantifies that of the block diagonal ones around their saddle point value. In the $n \to 0$ limit, and for any $\epsilon$, the above expression simplifies and yield
\begin{eqnarray}\label{eq:complete}
\fl
\left\langle \mathcal{N} \right\rangle = \rme^{N f_{n\to 0}(S_*)} \ee^{\frac{(1-\tau)}{2m^2}p^2} \left(1-\frac{\tau}{m^2}(p^2-r^2)\right)^{-\frac{1}{2}}\left(1-\frac{\tau}{m^2}(M_2^{-1})_{33}\right)^{-\frac{1}{2}}\,,
\end{eqnarray}
with
\begin{eqnarray}\label{eq:iverse33}
(M_2^{-1})_{33} = \frac{m^2\left((p^2-r^2)^2+m^2(p^2+r^2)\right)}{m^4+\tau(p^2-r^2)^2+(1+\tau)(p^2+r^2)m^2} \,.
\end{eqnarray}
We stress that while both $\det M_1$ and $\det M_2$ vanish at the $m<1$ saddle point when the limit $\epsilon$ is taken, the expression in Eq.~\eref{eq:complete} is free from any divergence. For $m > 1$, Eq.~\eref{eq:sol} implies in the limit where $\epsilon \to 0$,
\begin{eqnarray}
\left\langle \mathcal{N} \right\rangle = 1 \ .
\end{eqnarray}
For $m < 1$, we get from Eq.~\eref{eq:sol}
\begin{eqnarray}\label{eq:racine}
    \left\langle \mathcal{N} \right\rangle = \sqrt{\frac{1+\tau}{1-\tau}}\ee^{\frac{(1-\tau)(1-m^2)}{2}} \ee^{-N\left(\frac{1-m^2}{2}+\ln m \right)} \, .
\end{eqnarray}
Note that the small $O(1/N)$ contributions to the correlation matrix of the Jacobian in Eq.~\eref{eq:corrlJ} contribute to the pre-exponential factor of $\left\langle\mathcal{N}\right\rangle$ through the term $\ee^{\frac{(1-\tau)(1-m^2)}{2}}$, an innocuous contribution that is missing in the original work of \cite{fyodorov2016nonlinear}. This contribution can however be easily recovered from the results of \cite{fyodorov2016nonlinear} by performing the following substitution in Eq.~(14) of their work
\begin{eqnarray}
m \rightarrow m \left(1+\frac{1-\tau}{N}\right)^{-\frac{1}{2}} \,, \nonumber \\
\tau \rightarrow \tau \left(\frac{N+1-1/\tau}{N+1-\tau}\right) \,,
\end{eqnarray}
as suggested by Eq.~\eref{eq:corrlJ}. Even then, we note the existence of a $\sqrt{2}$ discrepancy between the result Eq.~\eref{eq:racine} and the rigorous random matrix theory result of \cite{fyodorov2016nonlinear}. It is legitimate to wonder whether our ansatz could have missed other saddle points the contributions of which could restore the missing $\sqrt{2}$ prefactor. We have reasons to believe that this is not the case, as we now discuss.

\subsection{Beyond the diagonal ansatz}\label{sec:ansatz}

The following discussion is inspired by the work of \cite{kamenev1999wigner} where fermionic replicas where used to derive the asymptotic of eigenvalue correlations in the Gaussian unitary ensemble of random matrices. We start by noticing that the integrand in Eq.~\eref{eq:integral2} is invariant under the action of the orthogonal group $\mathcal{O}_{n-1}$, \textit{i.e.} is invariant under the transformation
\begin{eqnarray}
S \rightarrow \left[ 
    \begin{array}{c|c} 
      O & 0 \\ 
      \hline 
      0 & O
    \end{array} 
    \right]S \left[ 
    \begin{array}{c|c} 
      O^T & 0 \\ 
      \hline 
      0 & O^T
    \end{array} 
    \right] \,.
\end{eqnarray}
with $O$ an orthogonal matrix $O \in \mathcal{O}_{n-1}$. This corresponds to a rotation of the matrices $P, Q$ and $R$ by the same orthogonal matrix $O$,
\begin{eqnarray}
& P \rightarrow O P O^T \, , \nonumber \\
& Q \rightarrow O Q O^T \, , \nonumber \\
& R \rightarrow O R O^T \, .
\end{eqnarray}
We next integrate over the orthogonal group by diagonalizing the matrix $P = O \Lambda O^T$ with $O \in \mathcal{O}_{n-1}$ and $\Lambda$ a diagonal matrix. By rotating accordingly the matrix $Q$ and $R$ we obtain
\begin{eqnarray}\label{eq:integral3}
\fl \left\langle I_{\epsilon}^{n-1} \right\rangle = C_{N,n} V_{n-1} \int \prod_{a = 1}^{n-1} d \lambda_a \prod_{a < b}^{n-1} | \lambda_a - \lambda_b | \int \prod_{a \leq b}^{n-1} d Q_{ab} \prod_{a, b}^{n-1} d R_{ab} \Theta(S > 0)\left(\det S\right)^{-n + \frac{1}{2}} \nonumber \\ \fl \exp{\left\{N \left[-\frac{\epsilon}{2}{\rm Tr}Q - \frac{\epsilon}{2}{\rm Tr}\Lambda + i {\rm Tr}R  - \frac{1}{2 m^2}\left( {\rm Tr}\left(Q\Lambda\right) + \tau {\rm Tr}R^2 + \tau\left({\rm Tr}R\right)^2\right) + (n-1) \right.\right.} \nonumber \\ \fl \left. \left. + \frac{1}{2}\ln \det S\right]\right\} \exp\left(-\frac{1-\tau}{2m^2}\left[{\rm Tr}\left(Q\Lambda\right) - \left({\rm Tr}R^2 + \left({\rm Tr}R\right)^2\right)\right]\right)\, ,
\end{eqnarray}
with $V_{n-1}$ the volume of the orthogonal group $\mathcal{O}_{n-1}$ and $\Lambda_{ab} = \lambda_a \delta_{ab}$. Following the notation introduced in Eq.~\eref{eq:inverse_notation} for the inverse matrix, the saddle point equations associated to Eq.~\eref{eq:integral3} write
\begin{eqnarray}\label{eq:saddle_point2}
      \tilde{\Lambda}_{aa} - \frac{1}{m^2}Q_{aa} - \epsilon = 0 \, , \nonumber \\
      \tilde{Q}_{ab} - \left(\frac{\lambda_a}{m^2} + \epsilon\right) \delta_{ab} = 0 \, , \nonumber \\
      \tilde{R}_{ab} + i \delta_{ab} - \frac{\tau}{ m^2}\left( \delta_{ab} {\rm Tr}R + R_{ba}\right) = 0 \, .
\end{eqnarray}
Equivalently, these equations can be rewritten as
\begin{eqnarray}
      \tilde{\Lambda}\Lambda + \frac{\tau}{m^2}\left(R^T\right)^2 - \left(i - \frac{\tau}{m^2}{\rm Tr}R\right)R^T  = \mathbbm{1} \, , \label{eq:sp1} \\
      \epsilon Q + \frac{\Lambda Q}{m^2} + \frac{\tau}{m^2}R^2 -  \left(i - \frac{\tau}{m^2}{\rm Tr}R\right)R = \mathbbm{1} \, ,  \label{eq:sp2}  \\
      \tilde{\Lambda}R + \frac{\tau}{m^2}R^T Q - \left(i - \frac{\tau}{m^2}{\rm Tr}R\right)Q = 0 \, , \label{eq:sp3}  \\
        \epsilon R^T + \frac{\Lambda R^T}{m^2}  + \frac{\tau}{m^2}R\Lambda - \left(i - \frac{\tau}{m^2}{\rm Tr}R\right)\Lambda = 0 \label{eq:sp4}  \, .
\end{eqnarray}
Combining Eqs.~\eref{eq:sp1} and \eref{eq:sp2} we get
\begin{eqnarray}
\tilde{\Lambda}\Lambda = \epsilon Q + \frac{Q\Lambda}{m^2} \Rightarrow Q_{aa} = \lambda_a \, .
\end{eqnarray}
Furthermore, from Eq.~\eref{eq:sp4} for all $a \neq b$
\begin{eqnarray}\label{eq:solR}
\left[ 
    \begin{array}{c|c} 
      \frac{\tau}{m^2}\lambda_b & \epsilon + \frac{\lambda_a}{m^2} \\ 
      \hline 
      \epsilon + \frac{\lambda_b}{m^2} & \frac{\tau}{m^2}\lambda_a
    \end{array} 
    \right] \left[ \begin{array}{c} 
      R_{ab} \\ 
      R_{ba}
    \end{array} 
    \right] = 0 \, ,
\end{eqnarray}
and for all $a$,
\begin{eqnarray}\label{eq:solRdiag}
\left(\epsilon + \frac{\lambda_a}{m^2}(1+\tau)\right)R_{aa} =  \left(i - \frac{\tau}{m^2}{\rm Tr}R\right)\lambda_a \, .
\end{eqnarray}
In the following, we assume that, as $\epsilon \to 0^+$, and in order to describe the multiple equilibria phase,  we can focus only on solutions of Eqs.~\eref{eq:sp1}-\eref{eq:sp4} that are such that $\lim_{\epsilon \to 0^+} \lambda_a \neq 0$. Under such an assumption, Eq.~\eref{eq:solR} implies $R_{ab} = 0$ for all $a \neq b$. Accordingly, we also obtain from Eqs.~\eref{eq:sp1}-\eref{eq:sp2} that $\tilde{\Lambda}_{ab} = Q_{ab} = 0$ for all $a \neq b$. All in all, the saddle point equations reduce to
\begin{eqnarray}
R_{aa} = \frac{i m^2}{1 + n\tau} \, ,
\end{eqnarray}
and 
\begin{eqnarray}
\lambda_a^2 = m^2\left(1 - \frac{m^2}{1+n\tau} - \frac{n\tau m^2}{(1+n\tau)^2}\right) \, .
\end{eqnarray}
As $n \to 0$ we recover for the matrix $R$ the solution obtained in Eq.~\eref{eq:sol} within the block identity ansatz. However, if we indeed get $\Lambda = Q$ as in Eq.~\eref{eq:sol}, the matrix $\Lambda$ need not be proportional to the identity as its eigenvalues are independently given by 
\begin{eqnarray}
\lambda_a = \pm \, m\sqrt{1-m^2}\, .
\end{eqnarray}
Note that all these are degenerate saddle points as for each of them the exponential weight of the integrand in Eq.~\eref{eq:integral3} is given in the limit $\epsilon \to 0^+$ and at finite $n$ by
\begin{eqnarray}
f^* & = \frac{n-1}{2}\left[1 + \frac{m^2 (-1+n\tau)}{(1+n\tau)^2} + \ln\left(\frac{m^2(1+n\tau(2-2m^2 + n\tau))}{(1+n\tau)^2}\right)\right] \, , \nonumber \\ 
 & \! \hspace{0.02cm} \!\underset{n\to 0}{=} \frac{m^2 - 1}{2} - \ln m \, .
\end{eqnarray}
The first line of the above equation gives the exponential weight of the negative integer moments of the determinant $\left\langle I_{\epsilon}^{n-1} \right\rangle$ for $n > 1$ and $\epsilon$ close to 0. Unlike the $n\to0$ result, the finite $n$ one displays an explicit dependence in the parameter $\tau$. As $n$ is sent to $0$, we assume that all these saddle points should be taken into account to get the order $O(1)$ corrections to the log-equivalent of $\left\langle {\mathcal N} \right\rangle$. We then parametrize each of them by $p$ defined as the number of eigenvalues $\lambda_a$ such that  $\lambda_a = - m\sqrt{1-m^2}$. By expanding around the different saddle points $S^*$ as $S = S^* +\hat{S}/\sqrt{N}$, we obtain
\begin{eqnarray}\label{eq:sumSP}
\fl  \left\langle I_{\epsilon}^{n-1} \right\rangle = \rme^{N f^*}  C_{N,n} V_{n-1} \, g \sum_{p=0}^{n-1} {n-1\choose p}
 \int \prod_a \frac{d \hat{\lambda}_a}{\sqrt{N}} \left(2m\sqrt{1-m^2}\right)^{p(n-1-p)} \prod_{a < b = 1}^p \frac{\left|\hat{\lambda}_a - \hat{\lambda}_b\right|}{\sqrt{N}} \nonumber \\ \fl \prod_{a < b = p+1}^{n-1} \frac{\left|\hat{\lambda}_a - \hat{\lambda}_b\right|}{\sqrt{N}} \int \prod_{a \leq b} \frac{d \hat{Q}_{ab}}{\sqrt{N}}\prod_{a, b} \frac{d \hat{R}_{ab}}{\sqrt{N}} \exp{\left(-\frac{1}{2}\hat{S}H_p \hat{S}\right)} \,,
\end{eqnarray}
with 
\begin{eqnarray}
g = \exp\left(\frac{(1-n) (1-\tau) \left(m^2 (-2 n \tau +n-1)+(n \tau +1)^2\right)}{(n \tau +1)^2}\right) \,,
\end{eqnarray}
and where $H_p$ is the Hessian of $f$ evaluated at any of the saddle points with $p$ negative $\lambda_a$. Collecting all powers of $N$ we obtain,
\begin{eqnarray}
\fl  \left\langle I_{\epsilon}^{n-1} \right\rangle = \rme^{N f^*} \left(\frac{1}{2}\right)^{n-1} \!\!\! \left(\frac{1}{2\pi}\right)^\frac{(n-1)(2n-1)}{2} \!\!\! \!\!\! V_{n-1} \, g \sum_{p=0}^{n-1} {n-1\choose p}  \left(2\sqrt{N}m\sqrt{1-m^2}\right)^{p(n-1-p)} \eta_p \, ,
\end{eqnarray}
with $\eta_p$ the value of the remaining $O(1)$ integrals in Eq.~\eref{eq:sumSP}. We stress that due to the Jacobian arising when going from the matrix $P$ to the vector of eigenvalues $\Lambda$, the different saddle points, while having the same exponential weight, come with different powers of $N$. Using the expression, 
\begin{eqnarray}
{n-1\choose p} = \frac{\Gamma(n)}{\Gamma(n-p)\Gamma(p+1)} \, ,
\end{eqnarray}
and the fact that $\Gamma(q)$ diverges for any negative integer $q$, the authors of \cite{kamenev1999wigner} proposed in a similar context to extend the sum over $p$ to infinity and use the resulting formula to carry on the analytical continuation to $n \to 0$. In this case, the different terms of the sum come with a contribution proportional to $N^{-\frac{p(p+1)}{2}}$ therefore suggesting that only the $p=0$ saddle point studied at depth in the previous section contributes to leading order in the limit $N \to \infty$.

\section{Outlook}
Our alternative replica-based calculation of the mean number of critical points in an $N$-dimensional dynamical system with a random Gaussian force, comprising both conservative and dissipative contributions, reproduces known results. There exists a transition between a regime with an exponential growth in $N$ of the number of critical points and a regime with a single critical point. This transition is driven by the amplitude of the random force.  In mathematical terms, and within our approach, the problem reduces to the calculation of an integral for which there exist two saddle points in the complex plane. The transition is explained by the fact that the integration contour can be deformed to catch either one or the other.
At a rather modest technical cost it does correctly reproduce the leading $N$ exponential growth of that number, within the simplest block-identity ansatz for our replica overlaps. Rather annoyingly, it seems that it misses an overall $\sqrt{2}$ prefactor (though, remarkably, it does catch the correct dependence on the parameter quantifying the lack of conservativeness of the random force field). We have shown that the calculation actually involves many other saddle solutions for the replica overlaps, which, we believe, and following the line of reasoning of \cite{kamenev1999wigner}, cannot be held accountable for this $\sqrt{2}$ discrepancy. One of the blind spots of the replica trick could be, when evaluating the contribution of the fluctuations, noncommuting $\epsilon\to 0$ and $N\to+\infty$ limits. This is strongly suggested by the fact that, for $m < 1$, our result is expressed as the (regular) ratio of two fluctuating determinants that become singular in the $\epsilon\to 0$ limit. It would of course be very interesting to precisely locate the mathematical hick-up. We note that, in a mathematically similar context \cite{fyodorov2014topology} (the computation of the large-deviation function of the ground state energy of a spin-glass model in a random magnetic field), a similar discrepancy at the pre-exponential level was identified between the results of the replica approach and exact random matrix theory calculations performed in the absence of external field. There, this discrepancy was also attributed to a non-commutativity of the large $N$ and the zero magnetic field limits and the appearance of non-Gaussian fluctuations. Among other research directions, we believe the approach presented here could be put to work out the moments of $\mathcal N$. The calculation would be more intricate, as the Kac-Rice formula would then involve averaging $f_i$'s and $\partial_j f_\ell$'s at different points in space.

\ack We thank Y. Fyodorov for very insightful discussions.

\section*{References}
\bibliographystyle{iopart-num}
\bibliography{biblio}

\end{document}